\documentclass[12pt]{article}
\pdfoutput=1
\usepackage{scicite}
\usepackage{times}
\usepackage{enumitem}

\usepackage{amsthm}
\usepackage{amsmath}

\usepackage{titlecaps}
\Addlcwords {a an the at by for in of on to up and as but or nor so yet vs with per} % Words not to capitalize in titles, "is" is captitalised because you always capitalise verbs
\usepackage[tiny]{titlesec}

\usepackage{float}
\usepackage{placeins}[section] % To control floating table/figure placement
\usepackage{CJK}
\usepackage{siunitx}
\usepackage{caption}
\usepackage{subcaption}
\usepackage{graphicx}

\usepackage{lineno}

\usepackage{xr}
\makeatletter
\newcommand*{\addFileDependency}[1]{% argument=file name and extension
  \typeout{(#1)}% latexmk will find this if $recorder=0 (however, in that case, it will ignore #1 if it is a .aux or .pdf file etc and it exists! if it doesn't exist, it will appear in the list of dependents regardless)
  \@addtofilelist{#1}% if you want it to appear in \listfiles, not really necessary and latexmk doesn't use this
  \IfFileExists{#1}{}{\typeout{No file #1.}}% latexmk will find this message if #1 doesn't exist (yet)
}
\makeatother

\newcommand*{\myexternaldocument}[1]{%
    \externaldocument{#1}%
    \addFileDependency{#1.tex}%
    \addFileDependency{#1.aux}%
}
\myexternaldocument{Supplement}

\usepackage{acro}
\acsetup{single} % Don't provide acronyms if term is only used once in the document.

\DeclareAcronym{L}{
    short=L,
    long=length of lymph node,
}

\DeclareAcronym{W}{
    short=W,
    long=width of lymph node,
}

\DeclareAcronym{H}{
    short=H,
    long=height of lymph node,
}

\DeclareAcronym{LN}{
    short=LN,
    long=lymph node,
}

\DeclareAcronym{EFPT}{
    short=EFPT,
    long=extreme first passage time,
}

\DeclareAcronym{LCMV}{
    short=LCMV,
    long=Lymphocytic choriomeningitis virus,
}

\DeclareAcronym{AIC}{
    short=AIC,
    long=Akaike's Information Criterion,
}

\DeclareAcronym{ABM}{
    short=ABM,
    long=agent based model,
}

\DeclareAcronym{DC}{
    short=DC,
    long=dendritic cell,
}
\DeclareAcronym{CRW}{
    short=CRW,
    long=Correlated Random Walk,
}

\DeclareAcronym{LMCRW}{
    short=LMCRW,
    long=log-mediated correlated random walk,
}

\DeclareAcronym{BM}{
    short=BM,
    long=Brownian Motion,
}
\DeclareAcronym{IFCT}{
    short=IFCT,
    long=Initial First Contact Time,
}

\usepackage{multirow}
\usepackage{url}
\usepackage{cleveref}
\crefname{theorem}{derivation}{derivations}
\Crefname{theorem}{Derivation}{Derivations}
\Crefname{assumption}{assumption}{assumptions}
\Crefname{assumption}{Assumption}{Assumptions}
\Crefname{corollary}{Prediction}{Predictions}

\usepackage{comment}
\usepackage{csquotes}
\usepackage{lettrine}

\newenvironment{sciabstract}{%
\begin{quote} \bf}
{\end{quote}}

\title{Bigger is Faster in the Adaptive Immune Response}

\author
{Jannatul Ferdous,$^{1\ast}$ G. Matthew Fricke,$^{1,2}$ Judy L. Cannon,$^{3}$ Melanie E. Moses $^{1,4,5}$ \\
\\
\normalsize{$^{1}$Department of Computer Science, The University of New Mexico, Albuquerque, USA}\\
\normalsize{$^{2}$Center for Advanced Research Computing, Albuquerque, USA}\\
\normalsize{$^{3}$Molecular Genetics and Microbiology, The University of New Mexico, Albuquerque, USA}\\
\normalsize{$^{4}$Department of Biology, The University of New Mexico, Albuquerque, USA}\\
\normalsize{$^{5}$Santa Fe Institute, Santa Fe, USA}\\
\\
\normalsize{$^\ast$To whom correspondence should be addressed; E-mail:  jannat@unm.edu.}\\
\normalsize{$^\dagger$These authors contributed equally to this work.}
}

\date{}

\begin{document} 
\pagenumbering{arabic}
\pagestyle{plain}

\baselineskip24pt


\maketitle


\begin{sciabstract}
Zoonotic pathogens represent a growing global risk, yet the speed of adaptive immune activation across mammalian species remains poorly understood. Despite orders-of-magnitude differences in size and metabolic rate, we show that the time to initiate adaptive immunity is remarkably consistent across species. To understand this invariance, we analyse empirical data showing how the numbers and sizes of lymph nodes scale with body mass, finding that larger animals have both more and larger lymph nodes. Using scaling theory and our mathematical model, we show that larger lymph nodes enable faster search times, conferring an advantage to larger animals that otherwise face slower biological times. This enables mammals to maintain, or even accelerate, the time to initiate the adaptive immune response as body size increases. We validate our analysis in simulations and compare to empirical data. 
\end{sciabstract}

\lettrine[lines=2]{M}{}ammal body masses range over 8 orders of magnitude, from the \SI{2}{\gram} bumblebee bat to the \SI{150000}{\kilo\gram} blue whale. Most biological processes slow with increasing body size, following a quarter-power scaling law \cite{kleiber1947body, west1997general, banavar2010general}. While the cause of quarter-power scaling is debated \cite{calder1996size, lindstedt1981body, mordenti1986man}, empirical observations consistently show that smaller mammals have faster physiology and life history, and larger mammals have slower rates over longer times \cite{charnov2007lifetime, west2000scaling, peters1986ecological, savage2004predominance}. For example, humans who are \si{2500} times larger than mice, are predicted to have heart rates, breathing rates, and gestation times that are 7 times slower than mice; actual values are 7 to 14 times slower, within a factor of 2 of the prediction \cite{charnov2007lifetime, clarke1976rhythm, meijer2006effect}. 

\begin{figure}
\centering 
\includegraphics[width = \columnwidth]{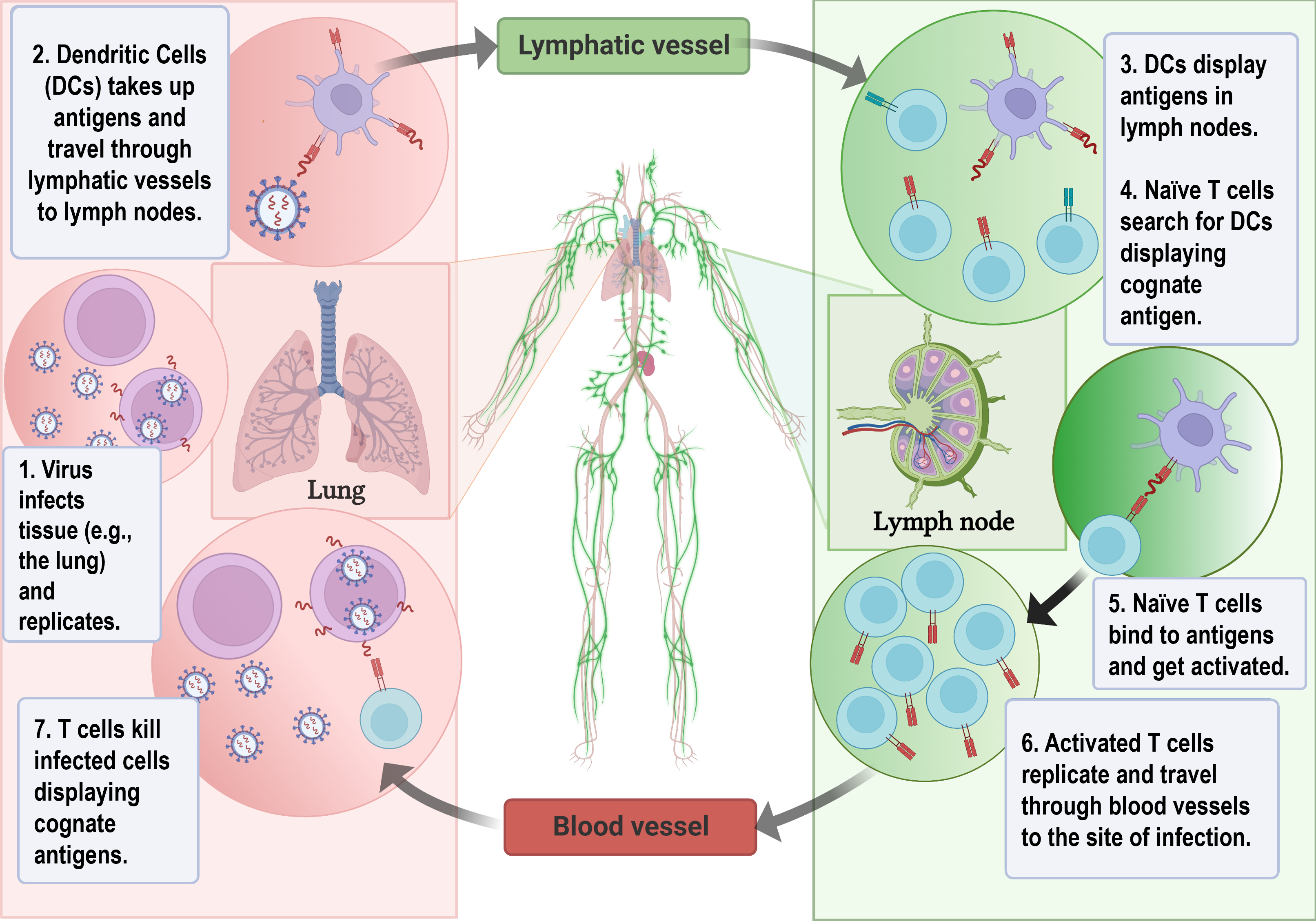}
\caption{{\bf \titlecap{Simplified schematic of T cell activation by Dendritic Cells.}}  %Figure shows a viral infection, but the response to bacterial infection is similar. 
1) A pathogen infects tissue, in this case for illustration, the lung. 2) \Acp{DC} deliver the captured antigens from tissue through lymphatic vessels to draining \acp{LN}. 3) \Acp{DC} display antigens in the \ac{LN}. 4) Na\"ive T cells search for cognate antigens presented on the surface of \acp{DC}. 5) T cell receptors recognize the cognate antigens upon encountering the antigen-bearing \ac{DC} and get activated upon receiving the activation signal from antigen-bearing DC. 6) Activated T cells proliferate exponentially, and CD8+ T cells transform into cytotoxic T cells (CTLs) that travel through the bloodstream to the inflamed, infected area. 7) CTLs kill the infected cells that display cognate antigens. We model the timing of search and activation in steps 4 and 5, where the adaptive immune response is initiated; the timing of this process depends on LN size.}
\label{fig:immuneResponse}     
\end{figure}

Despite the orders of magnitude increase in size and the slower metabolic rate of humans, the initial detection of the primary T cell response time in humans is indistinguishable from that of mice (\Cref{tab:immune_response_times}). Large animals clearly require that the immune response remain fast enough to counter exponentially growing pathogens. However, the mechanisms that allow larger mammals to respond as quickly as smaller, metabolically faster, ones remain unclear. The immune response proceeds through a sequence of interdependent steps, each reliant on the preceding one (\Cref{fig:immuneResponse}). Efficient scaling requires that none of these steps becomes a bottleneck. 

\Acp{LN} play a central role in this process. \Acp{LN} are the organs in which antigens indicative of infection are first recognized by T cells capable of mounting a virus-specific defense. We propose that the scaling rules governing the number and size of \acp{LN} help explain why two critical steps,  transport of antigens to \acp{LN} (step 2) and T cell contact with antigens carried by DCs within \ac{LN} (steps 3-5), remain fast across body sizes. 

\textbf{Our analysis considers a simplified model of immune response.} We primarily focus on Figure 1, steps 3 - 5 within \acp{LN}. We focus on generic lymph‐node search dynamics without distinguishing CD4$^+$/CD8$^+$ specific mechanisms.
We simplify the adaptive immune response to generalized steps beginning with infection at peripheral tissue sites (Fig. 1, step 1) where pathogens can establish and replicate.  To initiate the adaptive immune response, dendritic cells (DCs) in the tissues activate via Pattern Recognition Receptor signaling to ingest and process antigens produced by pathogens; they then upregulate migration receptors such as CCR7, and migrate via lymphatic vessels to draining LNs (step 2). While different pathogens activate different pathways and subsets of immune cells, such factors are not known to substantially affect the timing of these steps. DCs in LN display antigens (step 3) and na\"ive T cells move through LNs in search of cognate antigen-bearing DCs that they can bind to (step 4). For simplicity we consider CD8+ T cells that bind and activate and stimulate (step 5) and then migrate through the blood to the site of infection (step 6), where they kill infected cells displaying cognate antigens (step 7).  
This simplification of the very complex immune response focuses on CD8+ T cell activation that underlies anti-viral adaptive immunity, and not binding differences to MHC classes governing CD4+ T cell activation or high affinity antibody generation. We focus on the timing of T cell contact with DCs in LNs because this is the initiating event leading to other downstream adaptive immune responses.

\textbf{Scaling Context:} A well-established scaling relationship is that organ size typically scales linearly across animals. For example, the heart, liver, and kidney are 1000 times larger in animals weighing 1000 times more \cite{calder1996size}. We show that \acp{LN} deviate from this pattern and argue that the non-linear allocation of \ac{LN} size and number contributes to the invariance of the immune system response time.  We establish scaling relationships for how \ac{LN} volume, and \ac{DC} and T cell populations scale with body mass, and then we 
analyze how those scaling relationships determine how quickly the first T cells come into contact with \acp{DC} carrying cognate antigens in \acp{LN}.

\begin{figure}[!hbtp]
  \centering
  \includegraphics[width=\columnwidth]{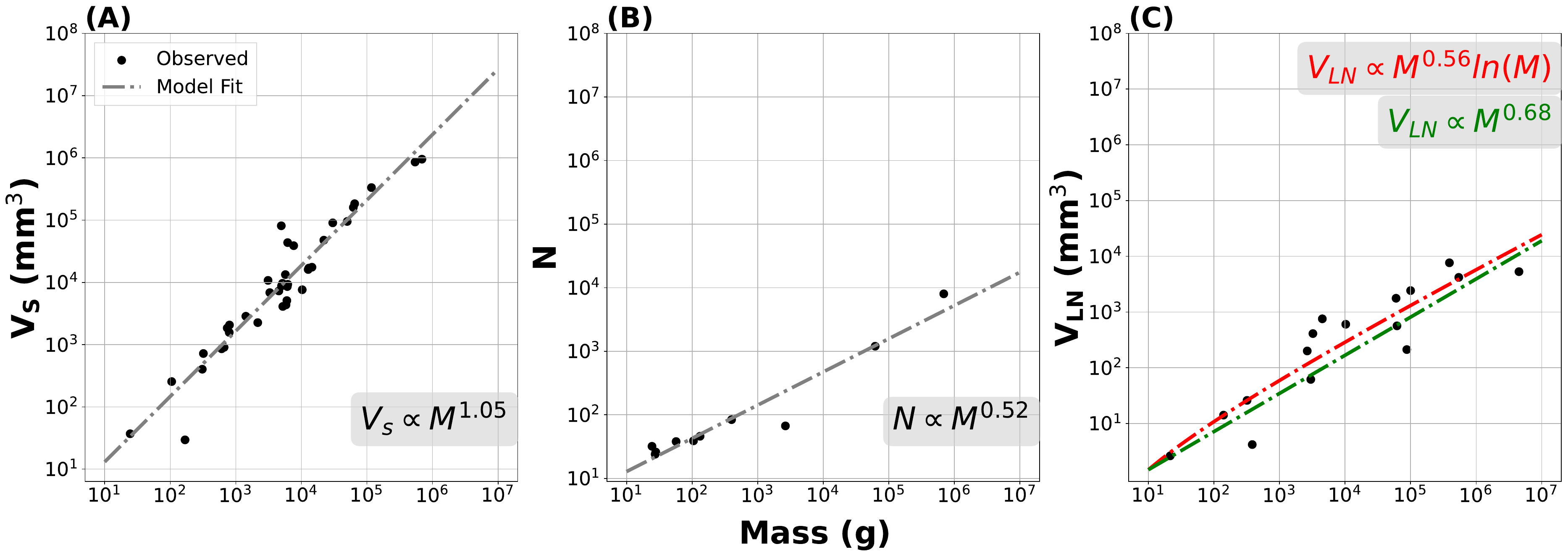}

  \caption{{\bf \titlecap{Lymphoid organ scaling with mass.}} Each data point represents a species. Both axes are on a log scale. The dashed lines show the reported regression fits. {\bf \titlecap{(A) Spleen volume}} of 38 species is best fit by the regression $cM^\mu$ with $c=\num{1.1609}$ and exponent $\hat\mu=1.05$ (95\% CI [\num{0.95}, \num{1.19}]). {\bf \titlecap{(B) Number of lymph nodes}} for 10 species is best fit by $cM^\mu$ with $c=\num{3.98}$ and $\hat\mu=\num{0.523}$ (95\% CI[\num{0.40}, \num{0.64}]). {\bf \titlecap{(C) Lymph node volume}} for 16 species is equally well fit in two ways: (1) a theoretically motivated fit including a logarithmic term $c_1M^\mu\ln(c_2M)$ (red line) with $c_1=1$, $c_2=1$, and $\hat\mu=\num{0.56}$ (95\% CI [\num{0.18}, \num{0.94}]) and by (2) by a simpler scaling fit, $c_1M^\mu$ (green line) with $c_1=1$, and $\hat\mu=\num{0.68}$ (95\% CI [\num{0.51}, \num{0.85}]). The p-value of the exponents is significant at the 0.01 level.}
  
  \label{fig:ScalingAll}
\end{figure}

The affect of \ac{LN} scaling on immune response has been studied previously \cite{perelson2006complexity,perelson2009scaling, banerjee2010scale, althaus2015mice}. Of particular relevance here, Perelson and Wiegel \cite{perelson2009scaling} theorized that if the benefits of larger \ac{LN} size and number were equally important and the total volume of \acp{LN} scales linearly with body mass, then \ac{LN} size and number should scale with the square root of body mass ($M^\frac{1}{2}$). 
%West et al. \cite{west1997general} famously, though not uncontroversially \cite{glazier2005beyond},  derived $M^\frac{3}{4}$ scaling laws from the properties of fractal networks, such as the cardiovascular network.
%\textcolor{red}{This paragraph needs to move to Results}\Cref{fig:ScalingAll}B shows empirical data on \ac{LN} number for 10 species that supports Perelson and Weigel's prediction. \Cref{fig:ScalingAll}C and D show \ac{LN} volume data for 16 species. The scaling of \ac{LN} volume is slightly steeper than Perelson's prediction, even when including a logarithmic growth term that Perelson hypothesizes is necessary to accommodate greater T cell diversity in larger animals. The data are also consistent with a $M^\frac{3}{4}$ scaling with body mass without the log term. % and equally well Perelson's scaling $\ln(cM)M^\frac{1}{2}$ scaling ($R^2=0.82$).
%, even as individual lymph nodes become proportionally smaller and less numerous.
For comparison, we show that the spleen, like most organs, scales approximately linearly with body mass in \Cref{fig:ScalingAll}A.%, with an exponent of 1.05 (indistinguishable from 1). 

We relate the speed of antigen detection in the \ac{LN} to theoretically predicted and empirically observed volume scaling observations with the formula, $M^{v-(t+d)}$, where $M$ is mass, and $v$, $t$, and $d$ are the scaling exponents relating \ac{LN} volume, the number of T cells, and the number of \acp{DC}, respectively, to $M$. We define \ac{IFCT} as the time it takes for the very first na\"ive T cell to come into contact with a cognate antigen in a \ac{LN}, and show that when larger \ac{LN} have more T cells and \acp{DC}, \ac{IFCT} is faster.  

%[MM: I think some of these numbers are a little wrong and we can make the .2 percent to 4 percent point later, after our main points]. Both \ac{LN} number and \ac{LN} volume scale sublinearly with body mass, a constraint imposed by the total lymphoid tissue volume, which itself follows $M^{1.1}$.  To illustrate, a lymph node mass fraction of 0.2\% in a mouse would translate to a fourfold higher fraction, approximately 4\%, in an elephant.  If \ac{LN} size were held constant at the mouse scale while \ac{LN} number alone increased, individual nodes would lack sufficient capacity for the expanding T-cell repertoire (which grows as $\ln(cM)$), and the likelihood of a cognate T-cell encountering its antigen-presenting dendritic cell within a node would diminish catastrophically.  Consequently, larger mammals must co-scale both \ac{LN} number and volume to sustain rapid adaptive immune surveillance.  

The benefit of more \acp{LN} is clear because a higher density of \acp{LN} reduces the average distance between potential infection sites and the closest \ac{LN} and therefore the antigen transport time (\Cref{fig:immuneResponse}, step 2) \cite{banerjee2010scale}. However, the benefit of larger \acp{LN} was previously not obvious, especially since Perelson and Weigel \cite{perelson2009scaling} predict that \emph{typical} search times should be independent of the \ac{LN} volume. That is, if the density of cells is constant, then a typical T cell or B cell would find a fixed target in the same amount of time, for any \ac{LN} volume. %(\Cref{fig:immuneResponse} steps 3-5). 

%Contrary to \Cref{fig:ScalingAll}C and D, if there is no benefit to larger \acp{LN} 

However, if there were no benefit to larger \ac{LN} volumes, it would be optimal to simply have as many \acp{LN} as possible to minimize the time for \acp{DC} to transport antigen to the \ac{LN} (\Cref{fig:immuneResponse}, step 2). Empirical data show that both the number and size of \acp{LN} increase with body mass, but sublinearly with exponents close to 1/2, but with the volume exponent slightly higher than the number exponent. %Mammals disproportionately allocate lymphoid tissue volume to larger \acp{LN} rather than to more numerous ones as body size increases. 
One explanation for this was proposed in \cite{perelson2009scaling}: larger, and generally longer-lived mammals encounter a greater diversity of pathogens, and therefore need larger \acp{LN} to maintain a greater diversity of immune cells. \Cref{eq:IFCT_scaling} suggests a complementary advantage to larger \acp{LN}: larger \acp{LN} hold more copies of T cells cognate to particular antigens, resulting in
reduced \ac{IFCT}.

In previously published work \cite{doi:10.1089/cmb.2023.0296}, we present a mathematical model that predicts \ac{IFCT} between searchers and targets distributed at random in a volume. We explored how the number of searchers, the distribution of searchers and targets, and the initial distances between searchers and targets affect \ac{IFCT}. Here, we build on those models to make a mathematical prediction for \ac{IFCT} scaling in \acp{LN} and test it in simulations. We show that the time to first T cell contact with a \ac{DC} is invariant with body mass given a constant number of \ac{DC}, as long as T cell density within \acp{LN} is fixed. Further, \ac{IFCT} decreases when both T cell and \ac{DC} density are constant. %scale with lymph node volume, and increases when neither T cells nor \acp{DC} scale, due to dilution effects in larger lymph node volumes. 
%Crucially, this reduction in \ac{IFCT} does not require increased cell density. Rather,
One key assumption is that there is a constant density of T cells in \acp{LN} (and this holds for any scaling of \ac{LN} volume with $M$), so larger \acp{LN} contain more T cells in absolute terms. Thus, in larger \acp{LN},  the probability of a T cell–DC pair encountering each other increases, reducing expected times to initiate the adaptive immune response.

\textbf{We base our analysis on the following simplifying Assumptions:} %about how naive T cells search for antigen in lymph nodes. The most important assumptions are listed here with more detailed justification provided in the methods section. In particular, we assume that
\begin{enumerate}
    \item T cell density is constant; it does not vary systematically with LN volume or animal mass. 
    \item T cells and \acp{DC} are uniformly distributed within the T‐cell zones of \acp{LN}.
    \item Cell‐cell encounters follow a memoryless exponential waiting‐time distribution.
    \item T cells move by unbiased diffusion in the LN. %so that the mean search time for a T cell to encounter a target DC scales linearly with lymph‐node volume. 
    \item Scaling exponents are not sensitive to prefactors that might represent details of particular subtypes of immune cells or pathogens, movement patterns of T cells, geometrical shapes of \acp{LN} or how cells enter \acp{LN}, nor to the noise inherent in data collected from published literature. We simplify the complex immune response in favor of a more general model. 
    \item We assume the density of \acp{DC} in \ac{LN} can vary. We consider two bounding cases: a) the number of \acp{DC} is constant or b) the density of \acp{DC} is constant with respect to \ac{LN} volume. 
    \item We do not know what fraction of na\"ive T cells are cognate to antigens produced from any particular pathogen. We model two alternative assumptions: a) the density of \emph{cognate} T cells remains constant across \acp{LN} (proportional to the density of all T cells), or b) increased diversity of T cells dilutes the density of cognate T cells by a logarithmic factor.
%(See Supplement \Cref{subsec:ABM} for more details as well as caveats and limitations.)
    \end{enumerate}

Numerous agent‐based and ODE models have explored how T cells scan antigen‐bearing \acp{DC}, examining effects of motility, affinity, and spatial organisation \cite{moreau2016virtual,bogle2010agent,bogle2012lattice,riggs2008comparison,gong2015quantifying,celli2012many,beltman2009analysing,graw2009investigating,donovan2012t,regoes2007estimation,textor2014random,fricke2016persistence,moses2021spatially}. These studies demonstrate that individual T‐cell–DC contacts can be prolonged and that not every T cell must engage for immunity to initiate. However, none has treated the \emph{very first} cognate encounter (which we call \ac{IFCT}) as a biologically meaningful threshold marking the true onset of the adaptive cascade. 

\ac{IFCT} marks the moment when a single cognate T\,cell first encounters its antigen-presenting \ac{DC}. Any delay directly postpones the immune peak because this time determines the earliest possible start of exponential clonal expansion. Hence, IFCT sets a lower bound on how fast the peak can be reached. Unlike peak response timing, IFCT depends solely on search dynamics within \acp{LN}, making it a key measure for understanding how the sizes and numbers of \acp{LN} can compensate for slower physiology to preserve rapid detection. 

Here, we show, both analytically and in agent‐based simulations, that \ac{IFCT} depends on the number of T cells and DCs involved in the search, and given more searchers in larger \acp{LN}, \ac{IFCT} is equally fast or faster in larger mammals.

\section*{\titlecap{Time to initiate the adaptive immune response is the same in humans and mice}}
We first establish that the timing of the first detectable adaptive immune response is similar in humans and mice. Data on the timing of immune response are available for multiple pathogens in mice and humans because mice are the predominant model organism in immunological research, and human data are of direct clinical relevance. %We also include data from macaques and pigs, but these are relatively understudied, and any cross-species comparisons involving them should be interpreted with greater caution. 
\Cref{tab:immune_response_times} shows that for a range of viral and bacterial pathogens, newly activated T cells are first detected in \acp{LN} or tissues in both species within 4–10 days, with a typical detection time of 6 days, following the activation of na\"ive T cells that had not previously encountered these antigens. The time to detect activated T cells reflects the time for cells to move, activate, and proliferate (Figure 1, steps 1-6 if T cells are detected in LN, or steps 1-7 if detected in infected tissues). In the rest of this paper, we focus primarily on a subset of these steps, T cell search for cognate antigen-presenting \acp{DC} in \acp{LN} (Figure 1, steps 3-5).

We note that the first detection of activated T cell populations is distinct from the peak T cell concentrations that are often measured in blood. It can take additional time to reach the peak after initial activation, particularly in larger animals. For example, peak T cell concentrations are observed in 5-10 days in mice \cite{zhuang2021mapping, miao2010quantifying} and 14-28 days in macaques and humans \cite{mattoo2022t, sette2021adaptive}.

\begin{table}
\begin{center}
\caption{{\bf Time to Initial Detection of Activated T Cells in Mice and Humans} Data are rounded to the nearest day (d). $n$ is the number of published studies. Means are calculated from the midpoint of each reported range. Minimum and maximum values reflect the full span of reported values across all studies. (SARS2: Severe acute respiratory syndrome coronavirus 2; LCMV: Lymphocytic choriomeningitis virus; HSV: Herpes simplex virus; RSV: Respiratory syncytial virus).}
\label{tab:immune_response_times}
\begin{tabular}{ l l  l}
& \textit{M. musculus} (\SI{24}{\gram}) & \textit{H. sapiens} (\SI{62}{\kilo\gram}) \\
\hline
& Flu: \SI{5}{\day} \cite{owens1981dynamics,miao2010quantifying}    & Dengue: \SI{7}{\day} \cite{friberg2011cross} \\
& Flu: \SIrange{4}{6}{\day} \cite{keating2018potential}   & Flu: \SI{6}{\day} \cite{brown1985subclass}  \\
& Flu: \SIrange{5}{7}{\day} \cite{tamura2004defense}   & LCMV: \SIrange{4}{5}{\day} \cite{de2003different}  \\
& HSV: \SIrange{5}{7}{\day} \cite{brenner1994similar,coles2002progression}  & RSV: \SIrange{7}{10}{\day} \cite{guvenel2020epitope} \\
& LCMV: \SIrange{5}{7}{\day} \cite{homann2001differential}    & SARS2: \SI{4}{\day} \cite{mcaloon2020incubation,koblischke2020dynamics} \\
& SARS2: \SIrange{5}{10}{\day} \cite{zhuang2021mapping} & SARS2: \SI{6}{\day} \cite{jones2021estimating,rai2021incubation}  \\
& SARS2: \SI{7}{\day} \cite{schulien2021characterization}  & SARS2: \SI{7}{\day} \cite{lei2020antibody,iyer2020dynamics} \\
& Staph: \SI{6}{\day} \cite{schmaler2011t}  & Staph: \SI{7}{\day} \cite{brown2015memory,khatib2005time}  \\
& Staph: \SI{9}{\day} \cite{ridder2022kinetic}  &  \\
\hline
$n$ & 11  & 12\\
Mean & \SI{6}{\day} & \SI{6}{\day}  \\
Min & \SI{4}{\day}  & \SI{4}{\day}  \\
Max & \SI{10}{\day} & \SI{10}{\day} \\
\end{tabular}
\end{center}
\vspace{-4mm}
%\footnotesize{$^*$ Rhesus macaques (\textit{M. mulatta}, \SI{6.5}{\kilo\gram}) SARS2: \SIrange{7}{10}{\day} \cite{nelson2022mild}. Data from 6 rhesus macaques ranging in mass from \SI{3}{\kilo\gram} to \SI{10}{\kilo\gram}.}\\
%\footnotesize{$^*$ Pig (\textit{S. scrofa}, \SI{117.3}{\kilo\gram}) Flu: \SI{6}{\day} \cite{khatri2010swine,lange2009pathogenesis}. }
\end{table}

Quantitative measures of initial antigen‐specific na\"ive T cell activation are scarce outside of mice and humans due to experimental and ethical constraints. While other species, such as swine, non‐human primates, and certain rodents, are widely used in infectious‐disease studies, detailed early adaptive response kinetics are rarely reported; many studies do not collect data before day 7 or day 10 post infection. While we did not find relevant reports of multiple different infections for species other than humans and mice, we did find the initial detection of T cells in macaques for SARS-CoV-2 and pigs for influenza were 7-10 days \cite{nelson2022mild} and 6 days \cite{khatri2010swine,lange2009pathogenesis} respectively. These times are consistent with the data in \Cref{tab:immune_response_times}, motivating the question - why is the timing of the initial immune response so similar across animals that are so different in size?

\section*{Empirical Scaling of Spleen Size}
\Cref{fig:ScalingAll}(A) shows that spleen volume \cite{mebius2005structure,lewis2019structure} scales approximately linearly with mass (also see Table S1).
%\Cref{tab:scaling_data}). 
A linear regression on log-log transformed data was used to derive an exponent, $\hat\mu$ of $1.05$ with 95\% CI [\num{0.95}, \num{1.19}] and with $R^2=\num{0.907}$. The data are consistent with the expectation of linear scaling of spleen size with $M$. The data are also consistent with an additional logarithmic increase ($M \ln(cM)$) (See Section S1.2).
%\Cref{sec:AlternateSpleenVol})
Such a nonlinear scaling could accommodate the predicted logarithmic increase in lymphocyte diversity with $M$ hypothesized in \cite{perelson2009scaling}. 

\section*{\titlecap{Empirical scaling of Lymph Node Number and Size}}

Lymph node numbers and volumes were estimated from healthy adult animals using standardized geometric approximations; details of data selection, geometric assumptions, and measurement methodologies are provided in Supplementary Section S1.1. While there is substantial heterogeneity in \ac{LN} volumes within each species (for example, human \acp{LN} range in diameter from $\SI{2}{\milli\meter}$ to $\SI{38}{\milli\meter}$), that variation is dominated by more than thousand-fold differences in LN volumes between the largest and smallest mammals in Supplementary Table S1. The scaling exponents we find by comparing across all species are consistent with the estimates we find comparing just between mice and humans, which are the best characterized species.

Wiegel and Perelson \cite{perelson2009scaling} propose \ac{LN} number and size scaling based on two key assumptions: first, maximizing \ac{LN} volume and number are equally important, and second, the total \ac{LN} volume scales approximately linearly, proportional to body mass $M$ (noting that scaling is also predicted to accommodate a logarithmic increase in T cell diversity with mass). Based on these assumptions, they predict that the volume of a typical \ac{LN}, $V_{\text{LN}}$, scales as follows, where $c$ is a constant,
\begin{equation}
\label{eq:LN_vol}
V_\text{LN} \propto M^\frac{1}{2}\ln(cM).
\end{equation}
The number of \acp{LN}, $N_{\text{LN}}$, scale as, 
\begin{equation}
\label{eq:LN_num}
N_\text{LN} \propto M^\frac{1}{2}
\end{equation}

\Cref{fig:ScalingAll}(B) shows the scaling for \ac{LN} number. The best fit for 10 species, with mass ranging from \SI{24}{\gram} mice to \SI{690} {\kilo\gram} horses, for $N_{LN} = cM^\mu$ is $c=\num{3.83}, \hat\mu=\num{0.523}$, 95\% CI [\num{0.40}, \num{0.64}], consistent with the Perelson-Weigel prediction of $\frac{1}{2}$ with $R^2$ = \num{0.9317}.

\Cref{fig:ScalingAll}(C) shows the scaling of \ac{LN} volume with mammal mass for 16 species from \SI{24}{\gram} mice to \SI{4500}{\kilo\gram} elephants. Regression of the form $V_{LN} = c_1M^\mu\ln(c_2M)$, produces, $c_1=\num{1}$, and $c_2=\num{1}$, $\hat\mu=\num{0.56}$ (95\% CI [\num{0.18}, \num{0.94}]) with $R^2$ = \num{0.8368}. While this is consistent with the predicted $\frac{1}{2}$ exponent, the inclusion of the log term allows flexibility in the fit, accommodating a very wide range of scaling exponents.  
Excluding the logarithmic term yields a higher exponent of  $\hat\mu=\num{0.68}$ (95\% CI [\num{0.51}, \num{0.86}]) (\Cref{fig:ScalingAll}C), with $R^2$ = \num{0.8341}. %consistent with other three-quarter power scalings commonly found in biology. 
The \ac{AIC} values for the models with and without logarithmic
terms are -23.42 and -23.26, respectively (See Section S1.3 %\Cref{subsec:statisticalAnalysis} 
for computation).

Interestingly, both scaling relationships suggest that total \ac{LN} volume (the number of \ac{LN} multiplied by typical \ac{LN} volume) scales superlinearly with body mass, as either $M^{1.08} ln(M)$ or $M^{1.20}$. Despite sublinear scaling of both \ac{LN} number and size, the total volume of \acp{LN} increases slightly superlinearly with body mass, implying that larger animals allocate a larger fraction of their body volume to lymphoid tissue. 

%$%$\hat\mu=$\num{0.809}, 95\% CI [\num{0.562}, \num{1.056}].

Since the data roughly align with the theoretical predictions given in \Cref{eq:LN_vol,eq:LN_num}, as well as the linear scaling of spleen volume, we can estimate human \ac{LN} volume, \ac{LN} number, and spleen volume relative to those of mice. The theoretical expectation is that \ac{LN} volume \si{350} times larger, \ac{LN} number 50 times larger, and spleen \si{2500} times larger in humans. Actual values from Table S1 are within a factor of two of these approximations. Given the more than three orders of magnitude difference in the sizes of humans and mice, predictions that are within a factor of two of empirical estimates are useful approximations, similar to the physiological scaling predictions of heart rates, breathing rates, and gestation times described in the introduction.
Given the similar fit for a simpler powerlaw equation $V_{LN} \propto M^\frac{2}{3}$, we analyze this scaling as well as the theoretically predicted $M^\frac{1}{2}ln(M)$ scaling. %\textcolor{red}{noting that our primary point does not depend on the exact exponent. Bigger \acp{LN} with more T cells and \acp{DC} lead to faster initiation of the immune response. The exponent determines the amount of speedup.} \textcolor{blue}{is that last phrase helpful or should we cut it?} JF: I understand what you are doing here. But I think putting it here will arise more questions about why do we need the scaling of the exponent, although we explainedn why later on. So I cut it.

\section*{\titlecap{Predicting Initial First Contact times}}

We derive a prediction for the time for the first T cell to find its cognate antigen-presenting \ac{DC} within a \ac{LN}, and then validate the prediction with our agent-based model (See Section S1.6
%\Cref{subsec:ABM} 
for detailed understanding of our agent-based model). We first consider a generic search problem between a population of T cells ($N_{\text{TC}}$) and a population of \acp{DC} ($N_{\text{DC}}$) that the searchers are looking for in a \ac{LN} volume ($V_{LN}$).
In Section S1.4,
%\Cref{subsec:derivation}, 
we derive an equation, Derivation 1,
%\Cref{thm:general_form_ExpVal},
for IFCT, represented by the variable $\tau_\text{init}$:
\begin{equation}
    \tau_\text{init} \propto \frac{\lambda}{N_{\rm TC} N_{\rm DC}}
     \label{eq:initial_first_contact_}
\end{equation}

where $\lambda$ is defined as the mean first-contact time in a volume between a single T cell and a single \ac{DC}. Celli et. al. \cite{celli2012many} showed that $\lambda$ scales linearly with volume ($\lambda \propto V_{LN}$) if the searcher and target are randomly placed and the searcher moves using Brownian motion (Definition 1).
%(\Cref{def:lambdaScaling}). 
Since we have the product of $N_{\text{TC}}$ and $N_{\text{DC}}$ in the denominator, search times decrease linearly with increases in both T cells and \acp{DC}. %\textcolor{red}{Jannatul replace $V$ with $V_{LN}$ in these 2 sections?} JF: Done.
%\MM{fixme. Let's not use ln(cM). Instead, address this after Eqn 4. We should probably use 3/4 scaling for our simulations since that's the best empirical fit.}

%\textcolor{red}{MM: here we need to shift from assuming 1/2 power scaling to Matthews generic scaling using t,v and d exponents. So I've commented out text about 1/2 power predictions so we talk about it after the next section that defines t v and d. I also deleted about a page and a half of text in an attempt to simplify. I think it works but checkme that I didn't cut anything important!}

% Using \Cref{thm:general_form_ExpVal} we  predict search times within \acp{LN}  assuming \Cref{eq:LN_vol}, i.e. that volume is proportional to $M^{\frac{1}{2}}\ln(cM)$. We also assume that the number of T cells of each clonal line is proportional to $M$, and therefore $N_{TC}$ is proportional to $M^\frac{1}{2}$ in each of $M^\frac{1}{2}$ \acp{LN}. 

\paragraph{Scaling of Initial First Contact Time ($\tau_{init}$)}

Assuming diffusive motion of cells within the \ac{LN}, let \ac{LN} volume scale as $V_{LN} \propto M^v$, the number of cognate T cells as $N_{\mathrm{TC}} \propto M^t$, and the number of cognate \acp{DC} as $N_{\mathrm{DC}} \propto M^d$. Then \ac{IFCT} scales as:

\begin{equation}
\label{eq:IFCT_scaling}
\tau_{\text{init}} \propto \frac{V_{LN}}{N_{\mathrm{TC}} N_{\mathrm{DC}}} \propto \frac{M^v}{M^t M^d} = M^{v - (t + d)}
\end{equation}
\Cref{eq:IFCT_scaling}, allows us to explore how different assumptions about how the mass scaling of \ac{LN} volumes, T cell numbers and DC numbers affect the time to initiate an immune response. \Cref{eq:IFCT_scaling} yields three scaling regimes:
\begin{itemize}[noitemsep]
  \item Case $i$: If $v < (t + d)$, then $\tau_{\text{init}}$ decreases with $M$. This represents faster scaling of $\tau_{\text{init}}$ in larger animals in \emph{systemic infections}. This occurs under Assumption 6a where $d=v$.
    \item Case $ii$: If $v=(t+d)$, then $\tau_{\text{init}}$ is invariant with $M$ (constant $\tau_{\text{init}}$). This represents constant numbers of \acp{DC} with respect to $M$ in \emph{localised infections} when $t=v$ and $d=0$. This occurs under Assumption 6b with Assumption 7a.
    \item Case $iii$: If $v>(t+d)$, then $\tau_{\text{init}}$ increases with $M$. This occurs with Assumptions 6b with Assumption 7b where $d=0$ and $t < v$ by a logarithmic factor.
\end{itemize}

We start with Assumption 7a that the density of cognate T cells are constant within \acp{LN}. In that case, it is the density of DCs that determine whether the scaling regime for $\tau_{\text{init}}$ is Case $i$ or $ii$. Following Assumption 6a or 6b, we consider two regimes as assumed bounds on what is biologically realistic. To analyze the systemic Case $i$, we assume the density of DCs carrying antigen in the \ac{LN} is constant ($d = v = t$, Assumption 6a). A constant density of DC's could be expected for a systemic infection, for example SARS-CoV-2 that infects some fraction of the lung, producing more total amounts of antigen, proportional to lung mass and body mass, $M$.
\begin{figure}[!hbtp]
\centering
\resizebox{0.95\linewidth}{!}{\includegraphics{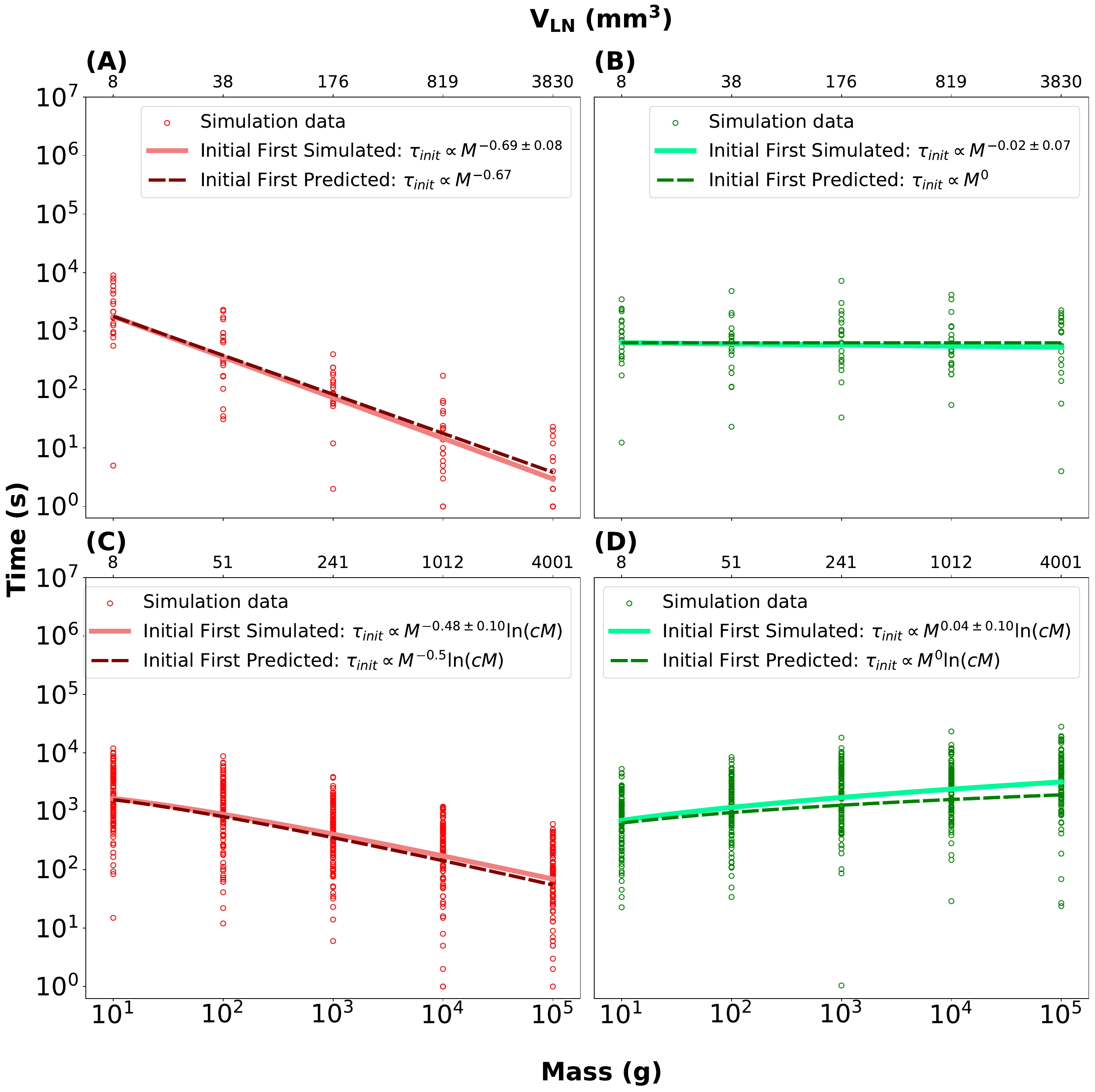}}
\caption{%
\textbf{Theoretical predictions vs.\ agent‐based simulations for initial T cell–DC contact time ($\tau_{init})$.} 
Simulation data (circles) are compared to predictions (dashed) and best‐fit curves (solid) under two volume hypotheses:
(A) Systemic infection assuming $V_{LN}\propto M^{2/3}$: predicted $\tau_{\rm init}\propto M^{-2/3}$ (dashed red), fitted exponent $\hat\mu=-0.69\pm0.08$ (solid red);  
(B) Localized infection assuming $V_{LN}\propto M^{2/3}$: predicted $\tau_{\rm init}\propto M^{0}$ (dashed green), fitted $\hat\mu=-0.02\pm0.07$ (solid green);  
(C) Systemic infection assuming $V_{LN}\propto M^{1/2}\ln(cM)$: predicted $\tau_{\rm init}\propto M^{-1/2}\ln(cM)$ (dashed red), fitted $\hat\mu=-0.48\pm0.10$ (solid red);  
(D) Localized infection assuming $V_{LN}\propto M^{1/2}\ln(cM)$: predicted $\tau_{\rm init}\propto M^{0}\ln(cM)$ (dashed green), fitted $\hat\mu=0.04\pm0.10$ (solid green).  
Each circle represents the simulation result of 20 simulation replicates in panels (A,B) and 100 replicates in panels (C,D) at each estimated LN volume. %\textcolor{red}{JF: is the 20 vs 100 still correct? If not please fix.} JF:Yes, correct%
}
\label{fig:predictions}
\end{figure} 

%\paragraph{Impact of scaling model on IFCT.}
%In both the Perelson and metabolic scaling regimes, 
%We first consider the systemic infection case where 

In this case, cognate T cell and DC counts scale linearly with \ac{LN} volume, i.e., both are constant density, so $t = d = v$. Thus, $\tau_{\text{init}} \propto M^{-v}$, and first contacts are faster in bigger animals. How much faster depends on the scaling exponent $v$.  We validate the prediction that $\tau_{\text{init}} \propto M^{-v}$ with agent based simulations shown in red in Figure 3A. We use $v=2/3$, as a simple approximation of the empirical data shown in Figure 2C (green line). The simulations show $\tau_{\text{init}}$ is faster, consistent with the scaling exponent of $-v$. Note that simulated volumes appear at the top of each panel and the corresponding animal mass is at the bottom of Figure 3. 

In the second case (Assumption 6b and Case $ii$) a localised infection might produce a fixed amount of infection, leading to a \emph {constant number} of antigen-bearing DCs in the \ac{LN}, so that the density of DC's decline with LN volume. Here we predict that $\tau_{\text{init}}$ is constant, as is validated in simulations in Figure 3B.  

In Figure 3C, we simulate LN volumes that scale according to Weigel and Perelson's theory, where $v=1/2$ multiplied by a ln(M) term that accommodates increased repertoire diversity in larger animals. Here we make Assumption 7b: LN size increases by a factor of $M^{1/2}$ multiplied by ln($M$) to accommodate increased diversity of T cell clonal lines. Thus, the density of particular T cells cognate to the antigens in the current infection increases by $M^{1/2}$ but decreases by ln($M$) (See Table S4). For systemic infections, Case $i$ again applies. Figure 3C validates that the predicted $\tau_{\text{init}}$ is
consistent with the predicted scaling, $M^{-1/2}ln(M)$. $\tau_{\text{init}}$ still decreases, but modified by a log factor generating a curvilinear fit. Finally, in Figure 3D, we show how $\tau_{\text{init}}$ scales under Assumptions 6b and 7b, a localised infection and the theoretical model. This produces a curvilinear logarithmic increase in $\tau_{\text{init}}$, following Case $iii$. 
% We next consider localised infections, something akin to a pinprick where the number of antigen-bearing DCs is constant with respect to LN volume. In equation 4, this means that $d=0$ and since we still assume T cell density is constant ($t=v$), then Equation 4 predicts $\tau_{\text{init}} \propto M^{0}$, or immune activation time is invariant with respect to mass. Figure 3 Panels B and D show simulations under the 3/4 scaling and Perelson-Weigel scaling assumptions. In the first case, $\tau_{\text{init}}$ is constant across LN volumes and animal masses. In the second case, where T cell density is lower by the log term, there is a small logarithmic increase in times.

\begin{table}[t]
\caption{{\bf \titlecap{Initial First Contact Times for Mice and Humans.}} We consider 2 cases each for the $\frac{2}{3}$ Scaling Model and the logarithmic Perelson Model: i) constant density of antigen-bearing \ac{DC} from a systemic infection and ii) a constant number of \ac{DC} in \ac{LN} for a localized infection. We estimate times for each case, considering that T cells move using Brownian motion.}
\centering
\begin{tabular}{ c | p{0.16\columnwidth} p{0.16\columnwidth} | p{0.16\columnwidth} p{0.16\columnwidth}}
\hline
& \multicolumn{2}{c|}{$\frac{2}{3}$ Scaling Model: $V_{\text LN} \propto M^{\frac{2}{3}}$} & \multicolumn{2}{c}{Perelson Model: $V_{\text LN} \propto M^{\frac{1}{2}\ln(cM)}$}\\ \hline % <-- 
DC scaling & $N_{\text DC} \propto M^{\frac{2}{3}}$ & $N_{\text DC} \propto M^0$ & $N_{\text DC} \propto M^{\frac{1}{2}}$ & $N_{\text DC} \propto M^0$\\ % <-- Combining two cells with alignment c| and content 12.
\hline
IFCT Prediction: & $M^{-\frac{2}{3}}$ & $M^{0}$ & $M^{-\frac{1}{2}}\ln(cM)$ & $M^{0}\ln(cM)$ \\
\hline
\hline
\SI{24}{\gram} (Mouse) & \SI{16.174376}{\minute} & \SI{13.492115}{\minute}& \SI{22.360224}{\minute} & \SI{14.080673}{\minute}\\ 
\SI{1}{\kilo\gram} & \SI{1.211101}{\minute} & \SI{12.601615}{\minute}& \SI{6.764485}{\minute} & \SI{28.101503}{\minute}\\  
\SI{62}{\kilo\gram} (Human) & \SI{0.068800}{\minute} & \SI{11.684563}{\minute} & \SI{1.396281}{\minute} & \SI{48.996161}{\minute} \\
\hline
\hline
\end{tabular}
\vspace{-5mm} % Sledgehammer to resuce space after table
\label{tab:IFCT}
\end{table}

Figure 3 shows four possible relationships between $\tau_{init}$ and $M$, depending upon the density of cognate T cells and \acp{DC} in LN. The density of DCs is the dominant factor, changing predictions by a $M^{\frac{1}{2}}$ factor for systemic versus localized infections. The density of cognate T Cells has a logarithmic effect, depending on whether cognate T cell density is diluted by increased T cell diversity or held steady. \Cref{tab:IFCT} shows that for any of these cases, $\tau_{init}$ is very fast. The predicted $\tau_{init}$ is less than an hour for both humans and mice. %for the four cases in Figure 3: systemic and localized infections for both simple powerlaw scalings and Perelson models. The scaling relationships from both models predict $\tau_{init}$ of less than an hour\textcolor{red}{lets fix the IFCT vs tau init language - just use one of them instead of switching back and forth. I think I prefer tau init once we've introduced eqn 4.} 

%shows that both predicted and simulated $\tau_{init}$ for systemic and localized infections. 

The faster search times for systemic infections, and approximately invariant search times for localised infections in \Cref{fig:predictions}  arise because the time for the rare fortunate \emph{first} contact is expedited when more T cells are present. However, the advantage of a large population doesn't benefit the \emph{typical} T cell, as the last T cell-DC encounter takes far longer when there are more T cells.

In order to compare $\tau_{init}$ with the previous models that considered the typical time for an average T cell to contact its cognate DC,  we also model the median first-contact times ($\bar\tau$), the time for a typical T cell to contact its cognate antigen-bearing \ac{DC}. We model this for both systemic and local infection under Assumption 7b, following the approach outlined in \cite{perelson2009scaling} (See Figure S1 and Table S3).
%\Cref{fig:predictionsMFCT} and \Cref{tab:MFCT}). 
We calculated ($\bar\tau$) from the same simulations as ($\tau_{init}$), but running until all T cells had their first contact with a DC and then calculating the median of those times. See Section S1.4
%\Cref{subsec:derivation} 
for full derivation of $\tau_{init}$ and $\bar\tau$ for different cases and Table S2 for all the notations used in this work.

For systemic infections, assuming the theoretical model ($\tau_{init} \propto M^{-\frac{1}{2}}\ln(cM)$), then $\bar\tau$ scales logarithmically with body mass as $M^0\ln(cM)$ (See Supplement Prediction 4.1);
%(\Cref{thm:mean_constant_density_scaling}); 
for the local infection, where $\tau_{init} \propto M^0\ln(cM)$, then $\bar\tau$ scales as $M^{\frac{1}{2}}\ln(cM)$ (See Supplement Prediction 4.2).
%(\Cref{thm:mean_constant_number_scaling}).
In both cases, $\tau_{init}$ is a factor of $M^{v}$ faster than $\bar\tau$. 
$\tau_{init}$ takes less than one hour in humans and mice, but for systemic infection $\bar\tau$ takes 1 day for a mouse and over 3 days for a human (see Table S3).
%\Cref{tab:MFCT}). 
For localized infections, $\bar\tau$ also takes about 1 day for a mouse but takes 100 days for a human! In this case, the typical T cell contact ($\bar\tau$) happens long after the first T cells have begun to activate and exponentially replicate. In some cases, the typical T cell would not even activate until long after the infection is resolved.
$\bar{\tau}$ is 65 times slower than $\tau_{init}$ for a mouse, but 3700 times slower for a human. 
These calculations emphasize the critical role of timely first encounters in initiating an effective immune response, particularly in larger animals. We argue that $\tau_{init}$ is more consequential than $\bar{\tau}$ for the clonal expansion of effector T cells that exponentially replicate (Figure 1, step 6)  after activation and then travel to tissues to fight pathogens (Figure 1, step 7). 

%As we show in analysis in Supplemental XXX, the first T cells to encounter cognate angigen can undergo many rounds of exponential growth before the average T cell has had its first encounter.

%The difference between $\tau_{init}$ obtained from \ac{IFCT} vs. $\bar{\tau}$ has practical implications for immune system modeling. 

To further test our model, we reanalyze the empirical data presented in \cite{de2001recruitment} using the \ac{IFCT} model (see Section S2
%\Cref{sec:ModelValidation}
for details). The model in \cite{de2001recruitment}, assumes that contact between all T cells and \acp{DC} happens simultaneously at a time corresponding to $\bar{\tau}$. We implement this assumption in a Median model (see Section S2
%\Cref{sec:ModelValidation} 
for details) using the median contact time from our simulations. In Figure S2, we compare T cell population dynamics from our \ac{IFCT} model, which accounts for the time for each individual T cell to first contact its DC target, with data from \cite{de2001recruitment} over time and with data from Median Model. We parameterize our systemic infection model to reflect epitope‐specific T‐cell clone immunodominance and precursor frequency by using empirical activation times; NP118 and GP283 are LCMV‐derived epitopes presented via MHC class I. NP118 is immunodominant and we use the empirical data to estimate a shorter time to activation (6.8 hours) vs. GP283 which is subdominant with a longer estimated activation time (18.8 hours).
After 5-6 days post-infection, the Median model predicts peak T cell populations of $3.9\times10^7$ for NP118 and $1.2\times 10^6$ for GP283, whereas the \ac{IFCT} model predicts more than double these values at $9.1\times10^7$ and $3.4\times10^6$. Thus, in a mouse, a model using IFCT would predict more than twice the peak number of T cells compared to a Median model (See Table S6). 

We then scale the models up to estimate the peak T cell population in a larger volume.
%over 8 days in a simple model that focuses on the exponential growth of T cells upon contacting their target DC.
Compared to the Median model, the \ac{IFCT} model predicts peak T cell populations that are 40-times larger for the larger \acp{LN} (Figure S3). Thus, by accounting for the rare early first contact, we estimate far larger peak T cell populations, particularly in larger animals. Thus, not only is search faster in bigger \acp{LN}, but also, earlier contacts make vastly more T cells during the exponential growth phase.

\textbf{Several calculations provide context for our results.} First, we explain why empirical data are unlikely to distinguish whether the theoretical prediction ($V_{LN} \propto M^{1/2}ln(M)$) or the more parsimonious ($V_{LN}\propto M^{2/3}$) is a better fit to the data. Given the 5 orders of magnitude range of body masses between mice and elephants, ln($M$) is a factor of 10, equivalent to increasing the exponent by 0.2 (i.e., from 0.5 to 0.7). When $N_{LN}$ and $V_{LN}$ are multiplied to estimate the total volume of \acp{LN} in an animal, both generate a slight superlinear increase of ln(M) or $M^{0.2}$. Since \ac{LN} tissue occupies approximately 0.2\% of a mouse, we estimate \ac{LN} tissue occupies 2 - 4\% of an elephant. Further, the two models predict the largest animals have \ac{LN} volumes of 4000 mm$^3$ and 6000 mm$^3$, both similar to the estimates for camels and elephants in Table S1. Since both models are consistent with data, we simulate both the theoretical and simpler models in Figure 3. 

We find the theoretical model convincing because it accounts for increases in T cell diversity. However, it is not clear whether increased T cell diversity results in a logarithmic dilution in the density of cognate T cells (because there are more \emph{other} T cells that are not cognate to the antigens of a particular infection). Alternatively, the increased diversity may result in more clonal lines of T cells that are cognate to more antigens (potentially counteracting the dilution). Without data to distinguish between these alternatives, we use the theoretical model (with logarithmic reductions in density of cognate T cells) and the simple scaling model (where cognate T cell density is constant) to cover both of these cases. Table 2 shows that these different assumptions make little practical difference. Importantly,  $\tau_{init}$ is nearly as fast or faster in bigger \ac{LN}, and $\tau_{init}$ is much faster than $\bar\tau$, by a factor of $M^{v}$ in all of our modeled scenarios.

%predicts an animal of $10^5$ grams has average LN volumes of , both consistent with the range of LN volume estimates for animals of this size in Supplemental Table S1. 

%in order to understand why the theoretical logarithmic increase is so close to the simpler powerlaw increase, we note that the theory ($M^{1/2}ln(M)$) 

%\textcolor{blue}{In the Supplement \Cref{sec:ModelValidation}, we show a close correspondence between our model and empirically observed peak T cell population following absolute times to initiate the adaptive immune response (translated to fit data from the mouse spleen). 

%We also show that our estimated scaling exponents of $1/2$ for \ac{LN} number and between $1/2$ and $2/3$ for \ac{LN} volume are similar to the exponents relating these quantities between mice and humans, two species that differ in mass by 2500-fold and have far more detailed data than other species.
\section*{Discussion}
%\subsection*{Summary and main interpretation}
The time to initiate the adaptive immune response is similar in mice and humans despite three-orders of magnitude difference in their mass. %Available data is scarce for other species, limiting the conclusions we a can draw, but the greatest difference in mean response time between species was between pigs (2 studies) and macaques (1 study). Pig response time was 30\% faster than macaques even though pigs are 180\% more massive. 
This unusual mass invariance in initial adaptive immune response times is accompanied by an unusual scaling of the organs in which adaptive immunity is initiated. \ac{LN} number and volume both scale sublinearly with mammal mass ($M$), and the total volume of \acp{LN} scales slightly superlinearly with mass. The data are insufficient to differentiate whether total LN volume increases proportional to $M^1\ln(M)$ following Perelson and Weigel's earlier theoretical predictions, or with a simpler $M^{1.2}$ scaling equation.% (given the similar fit of using simpler regression models $V_\text{LN} \propto M^\frac{3}{4}$ and $N_\text{LN} \propto M^\frac{1}{2}$). 

%Thus, the average volume of a LN scales differently than the volume of the spleen and other organs that typically scale linearly with $M$. 
Theory \cite{perelson2009scaling}  predicts one-half exponents for the number and average volume of LN if scaling up LN size and LN number have equal benefits and the total LN volume is constrained to scale approximately linearly with animal mass. The empirical data are consistent with this theoretical prediction. An obvious benefit of having more \acp{LN} as animal mass increases is that the distance from a site of infection to the nearest \acp{LN} is reduced, reducing time to transport antigens to the \acp{LN} \cite{perelson2009scaling,banerjee2010scale}.
The analysis above shows a previously undescribed benefit of larger \acp{LN}: the search for antigen-bearing \acp{DC} happens in equal time or faster in larger \acp{LN} as long as T cell density is constant. Given that \ac{LN} are larger in larger mammals, T cells initiate adaptive immunity by contacting DCs in LNs nearly as fast or faster in larger animals. 

%and assuming constant density of T cells and \acp{DC} in \ac{LN} (as we do for systemic infections), we predict faster initial first contact times ($\tau_{init}$ in larger LN by a factor of $M^{-v}$ where $v$ is the exponent relating average LN volume to mass. We validated this prediction in our simulations (\Cref{fig:predictions}, panels A and C in red).

%At first glance, the scaling $\tau_{\text{init}} \propto M^{v - (t + d)}$ might suggest that faster \ac{IFCT} arises from increased cell density, since the exponent becomes more negative as $t + d$ increases. However, this interpretation is incomplete. 

%In this systemic case with constant density of T cells and \acp{DC} in \acp{LN} ($t = d = v$), then $\tau_{\text{init}}$ decreases with body size. %The key is not density, but the combinatorial increase in T–DC pairings. 
The logic is simple: if larger LNs contain more T cells and more \acp{DC} in absolute terms, then the first ``lucky" T cell that quickly contacts a DC will be faster. This speed up is not because T cells and DCs are closer \emph{on average}, but because very short distances to cognate DCs and very fortuitous movements toward DCs become more likely given larger populations. Since immune activation can be initiated by fortuitous early interactions, this enrichment in the tail of the spatial distance distribution dramatically reduces $\tau_{\text{init}}$. Thus, the first time any T cell contacts its cognate DC in a LN can occur in minutes, compared to days for the first time a typical T cell contacts its cognate DC ($\bar\tau$).

Our \ac{IFCT} model, parameterized to match empirical observations of T cell population growth in mouse spleens, shows a two-fold increase in T cell population compared to a model that only considers the median first contact time rather than the \ac{IFCT} (Section S2).
%(\Cref{sec:ModelValidation}).
When scaled to a larger spleen or LN (i.e., the spleen of a macaque or the LN of a cow) the peak number of T cells grows 40-fold larger when exponential growth starts at  $\tau_{init}$ rather than $\bar\tau$.  By accounting for exponential growth intiated by the earliest T cells to contact cognate antigen-bearing DCs, growth of T cell populations could fight pathogens more effectively than was previously modeled.

\paragraph{\textbf{Distributed Lymphatic versus Centralized Cardiovacular Networks:}} According to metabolic scaling theory, quarter‐power scaling relationships \cite{west1997general, banavar2010general} arise from systematic increases in transport time in larger animals, based on the assumption that resources flow outwards from a single, central source (the heart) through a fractal circulatory network. In contrast, the immune system relies on a distributed architecture of \acp{LN} \cite{ferdous2022modeling}, each acting as an independent “hub” where na\"ive lymphocytes and antigen‐bearing dendritic cells meet. Not only does this decentralization permit multiple, parallel activation sites, but individual nodes can also recruit immune cells from distant regions via the lymphatic and blood vessels during an active infection. Crucially, this alternative mode of resource distribution underpins our key finding: Bigger \acp{LN}, enable faster initial T‐cell–DC contacts as body size increases; this is because the absolute numbers of both T cells and DCs in each LN grow with node volume, speeding up $\tau_{init}$. This distributed architecture with more LN in larger animals means LN are, on average, closer to sites of infection than they would be in a centralized model. This reduces transport times to LN. They are also bigger, reducing search times within LN. Empirically, this allows the adaptive immune response to be initiated in constant time (Table 1) across animals that vary substantially in body size. This meets the evolutionary imperative to detect and control exponentially replicating pathogens in large and small mammals.

Here, we have highlighted one advantage that the distributed lymphatic network provides: balancing the speed of transport to \acp{LN} with many small \acp{LN} with the faster detection of antigen within a few large \acp{LN}. However, there are other constraints on \ac{LN} size and number. For example, LN must be big enough to hold a sufficient diversity of B and T cells and a sufficient number of exponentially growing activated B and T cells during an infection; both of these may vary with animal size.

\textbf{Caveats, Limitations and Open Questions:} While our modeling framework provides mechanistic insights into how \ac{LN} sizes and numbers enable rapid initial T cell–DC contacts across body sizes, readers are referred to Supplementary Section S3 for a detailed discussion of underlying assumptions, empirical uncertainties, as well as limitations and potential extensions of
our approach. A particularly noteworthy caveat is that some values in our datasets are difficult to measure precisely. The number of \acp{LN} in an animal may be under counted, and average \ac{LN} size may be overestimated if the smallest \ac{LN} are missed, the structure of the \acp{LN} often varies and these factors may particularly skew estimates in larger species.

We have simplified complex immunology and anatomy in favor of a simpler model. The size of the T cell zone relative to the measured \ac{LN} volume, the effective T cell–\ac{DC} encounter radius, the shape of \ac{LN} and the diffusion coefficient of T cells could all affect our IFCT models. Further, there is variation among the myriad subtypes of immune cells, receptors and molecules. However, we assume there is not substantial systematic variation across body sizes, so that they do not change how search times scale with mass. 
While we intend our analysis to be general enough to apply to both LN and spleens, the processes of transporting antigens to these tissues and the architecture of these tissues are different, so the absolute timing of first contacts may be different in spleens and \acp{LN}.

It remains an open question to fully explain how the scaling of LN size and number, the complex dynamics of replicating T cells \cite{de2013quantifying}, and the movement of both antigens and T cells into LN \cite{ferdous2022modeling} result in such similar times (6 days) to observe the first antigen-specific replicated T cells in both mice and humans in \Cref{tab:immune_response_times}. Such an explanation requires not just analysis of search times within \acp{LN} (Figure 1, steps 3 -5), but also times for DC to ingest and carry antigen to LN (steps 1 and 2), and subsequently for T cells to replicate, differentiate and travel back to infected tissue (steps 5-6). 

Despite these caveats, several observations support our estimated scaling exponents and conclusions. The \ac{LN} size and number scalings from the best studied species, mice and humans, are consistent with the $N_{LN} \propto M^{1/2}$ and the two formulations we consider for LN volume: $V_{LN} \propto M^{1/2}ln(M)$ and $V_{LN} \propto M^{2/3}$ (see Supplemental Section S1). 
We also tested model sensitivity to alternative non-Brownian movement patterns of T cells \cite{fricke2016persistence} and found that empirically observed persistent motion decreases cell contact times by a relatively small factor (see Table S5),
%\Cref{tab:IFCT_CRW}), 
but does not alter the scaling exponent, consistent with Assumption 5.

Our models are also consistent with the timing of empirical T cell dynamics in Supplemental Section S2. Finally, our main conclusions hold regardless of scaling exponents: $\tau_{init}$ is similar across body sizes for localized infections, faster in larger animals for systemic infections, and substantially faster than $\bar\tau$ for all infections.  %Our analysis of how early replication of the first T cells affects peak T cell numbers are consistent with data from \textcolor{red}{XXX, as we show in Supplemental Section XXX.}    
%\textcolor{red}{MM: add the new mouse human comparison as a reasonableness check. Maybe also something about the 1/2 log scaling vs 3/4 scaling.}

%MM: CUT this, I think we don't believe that those studies are comparable and it confuses the message. We note that some mouse studies using adoptive transfer and high-resolution sampling report antigen-specific T cell activation as early as 1–2 days post-infection. Such early detection methods cannot be ethically or practically applied in humans, where clinical sampling constraints limit detection to later time points. Therefore, while typical reports indicate 4–6 days to initial detection in both mice and humans, the underlying kinetics are likely comparable at finer temporal scales.

\textbf{Broader Implications:} Our analysis shows a benefit of large size that has not been previously appreciated. While bigger animals are usually slower, here we show that $\tau_{init}$ is faster in larger mammals. This makes sense intuitively because when there are more searchers, the first target is found faster. This phenomenon has been studied by physicists as extreme first passage times \cite{lawley2020universal}. In contrast to our findings of a linear speedup with size, previous extreme first passage time (EFPT) analyses find a much slower speedup that is only logarithmic with the number of searchers. The differences arise because EFPT considers an infinite number of searchers, all starting their search at the same physical location, with search trajectories that overlap. In contrast, in the \ac{LN} search problem, a finite number of dispersed searchers in the 3D volume of a \ac{LN} can be considered independent of each other, leading to the much greater (linear) advantage of large search populations that we identify here.

The different scaling properties of \ac{IFCT} and typical first contact times are particularly relevant when the first contact causes a cascade of downstream events. In the initiation of adaptive immunity, when cognate T cells contact \acp{DC}, the T cells replicate (\Cref{fig:immuneResponse}, step 5), and changes occur in the \ac{LN}, including slowing the egress of other T cells. Thus, the first contact changes the dynamics of subsequent searches. %The typical search time of the other searchers changes because the first contact has happened. 
Further, the exponential growth of T cells begins once the first contact is made. Subsequent T cell contacts can amplify the T cell response, but the initial first contact causes the first T cell replication that produces activated T cells to migrate to fight infection in tissues (\Cref{fig:immuneResponse}, step 7). The first arrival time of T cells in tissue is important in controlling exponentially growing pathogens, as has been shown in response to SARS-CoV-2 infection \cite{sette2021adaptive} and in our simulations of the timing of T cell response \cite{moses2021spatially}.

Understanding the different scaling properties of initial versus typical first contact times is also relevant for other immunological processes, for example, the B cell search for T cells in LN (modeled in \cite{perelson2009scaling}) and effector T cell search for infected cells in peripheral tissue (modeled in \cite{moses2021spatially}). The analysis here suggests that initial contacts may happen faster in larger animals with more immune cells, but last contacts might take longer \cite{lawley2023slowest}. Last contacts may be relevant for understanding the dynamics of final clearance of infections.  

This variation in immune response can affect the timing and duration of infection and infectiousness in animals of different sizes; this, in turn, can affect how diseases spread across animal communities \cite{downs2019scaling}.
The distinct scaling properties of first, typical, and last search times warrant further study in immunology and biology more broadly. The different times to achieve typical, first, and last search events affect any biological search that involves large numbers of searchers. For example, the first ant in a colony that finds food should similarly depend on colony size, and when that first event happens, communication of the food location changes the search times for the typical ant in the population \cite{flanagan2012quantifying,moses2019distributed}. Similarly, the first individual with a rare genetic mutation that confers some fitness advantage occurs faster in larger populations and then changes the downstream dynamics. Thus, we suggest that understanding how the timing of the initial first successful search depends on the number of searchers is an essential and previously neglected question in immunology and, more generally, in biology.
\bibliography{main}
\bibliographystyle{sciencemag}

\section*{Acknowledgements}We thank the University of New Mexico Center for Advanced Research Computing, supported in part by NSF, for high-performance computing resources, as well as the James S. McDonnell Foundation and NSF awards 2030037 and 2020247 for funding. Thanks to Chris Kempes, Sid Redner, Alan Friedman, and the Moses Computational Biology Lab for the helpful discussions and reviews of earlier versions of this manuscript.

\paragraph*{Author contributions:}JF wrote the software and performed computational experiments under the supervision of GMF and MEM. JF, GMF, and MEM wrote the manuscript and created the figures. JLC supervised JF on the selection of simulation parameters, active immune response times in humans and mice, and helped the authors place our results in context. JF compiled published lymphoid tissue measurements under the supervision of MEM. JF derived the mathematical expressions under the supervision of GMF and MEM. All conceptual content and implementations are entirely generated by the authors. The paper was drafted entirely by the authors, and we used ChatGPT in a few instances to clarify or shorten individual sentences. For example, we asked ChatGPT to rewrite this sentence:
"It has been unclear what drives a trade-off between \ac{LN} size and number"
Chatgpt produced: "The trade-off between \ac{LN} size and number remains unexplained."
This was done to edit approximately 5\% of the paper.
\paragraph*{Competing interests:}We declare we do not have any competing interests.
\paragraph*{Data and materials availability:}All data for spleen volume, \ac{LN} volume, and \ac{LN} number used in this paper are collected from the published literature and included in a Supplementary in Table S1. Raw data files for initial and median first contact time generated from our model are available online at Dryad: \url{https://datadryad.org/stash/share/IkjXOE0jiqZ_rc4vdGIeqTYp9e08eCJkz654SyeFQ0A}
All figures, except for Figure 1, are generated using Python 3 in a Jupyter Notebook and Adobe Illustrator. Figure 1 was generated using BioRENDER. The code for our agent-based model, mathematical analysis, and figure generation is available at \url{https://github.com/BCLab-UNM/BiggerIsFaster}.

%%%%%%%%%%%%%%%% SUPPLEMENT LIST %%%%%%%%%%%%%%%

% List the contents of your Supplementary Materials, including the numbers of any
% supplementary figures, tables, external data files etc. and any references that are
% cited only in the supplement. In this example, refs. 7-8 are cited only in the supplement.
% Fill out your numbers accordingly and delete any lines that aren't applicable.
% \subsection*{Supplementary materials}
% Methods\\
% Model Validation\\
% Caveats and Limitations\\
% Figs. S1 to S3\\
% Tables S1 to S6\\
% References \textit{(1-58)}\\ % automatically fills 

% For your review copy (i.e., the file you initially send in for
% evaluation), you can use the {figure} environment and the
% \includegraphics command to stream your figures into the text, placing
% all figures at the end.  For the final, revised manuscript for
% acceptance and production, however, PostScript or other graphics
% should not be streamed into your compliled file.  Instead, set
% captions as simple paragraphs (with a \noindent tag), setting them
% off from the rest of the text with a \clearpage as shown  below, and
% submit figures as separate files according to the Art Department's
% instructions.

\clearpage
\end{document}